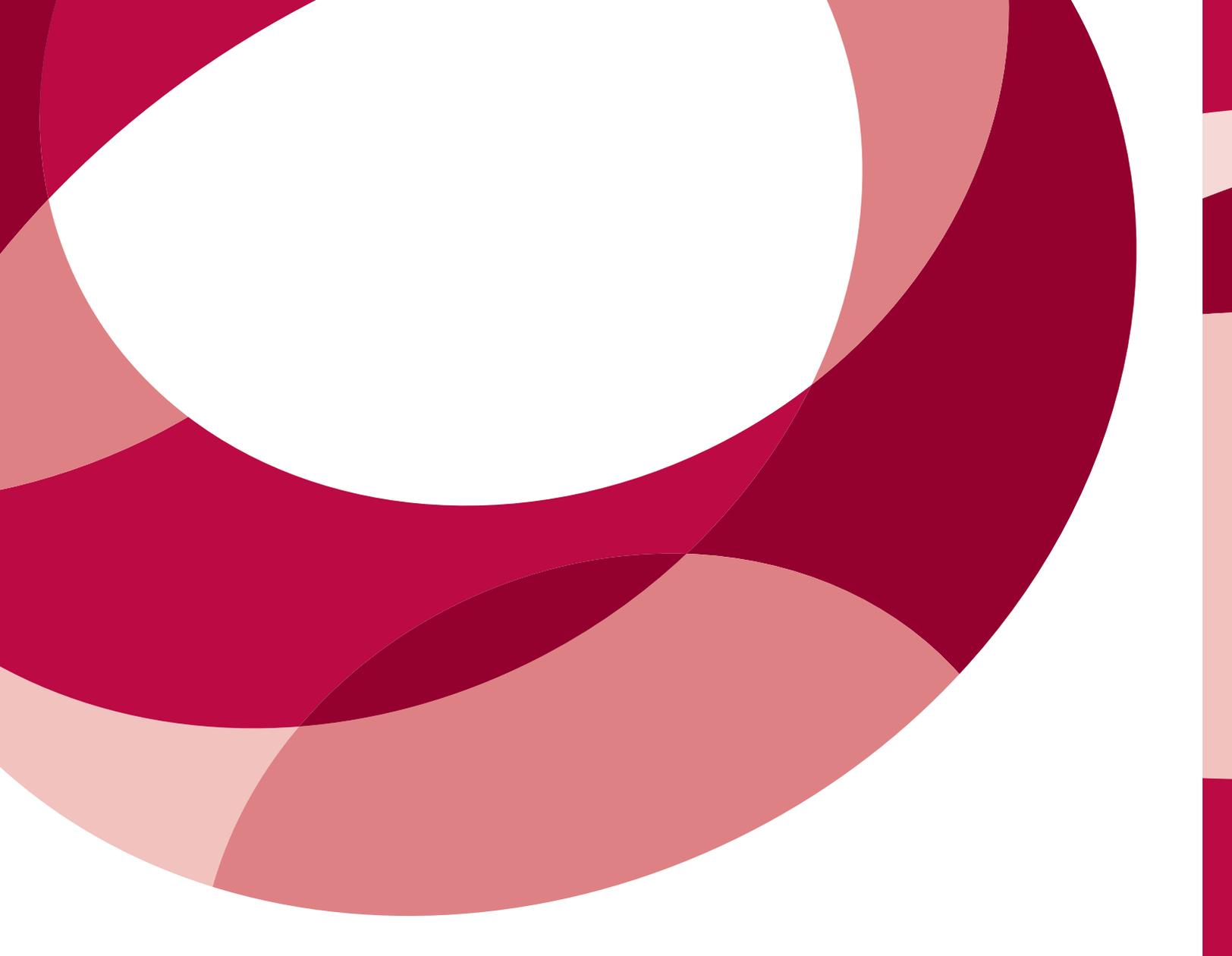

# Algorithmic and Economic Perspectives on Fairness

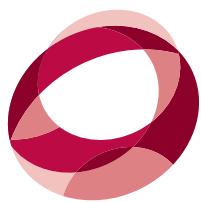

**CCC**
Computing Community Consortium
Catalyst

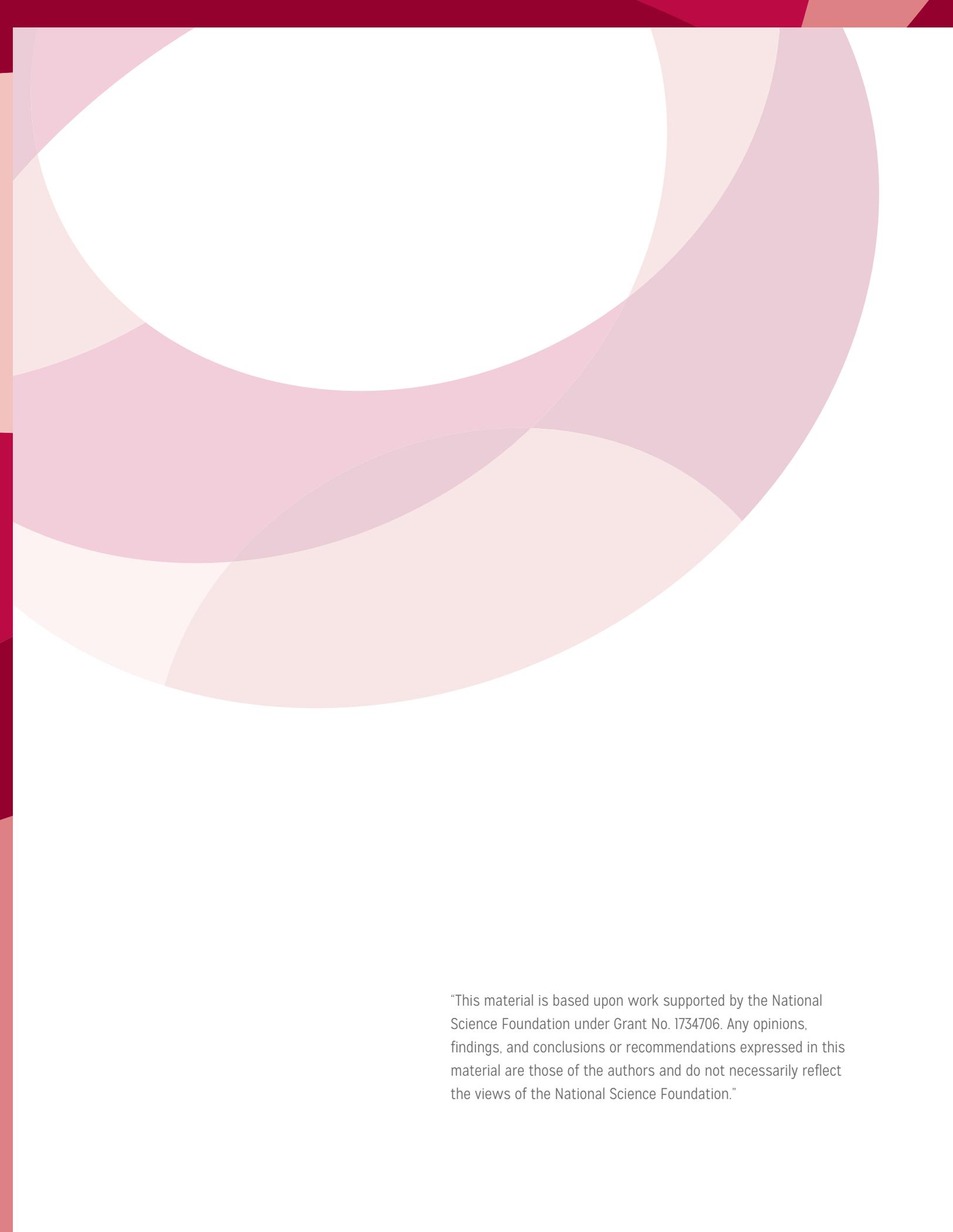
"This material is based upon work supported by the National Science Foundation under Grant No. 1734706. Any opinions, findings, and conclusions or recommendations expressed in this material are those of the authors and do not necessarily reflect the views of the National Science Foundation."

# Algorithmic and Economic Perspectives on Fairness

**A REPORT BASED ON A CCC FAIRNESS AND ACCOUNTABILITY TASK FORCE VISIONING WORKSHOP HELD MAY 22-23, 2019, AT HARVARD UNIVERSITY, CAMBRIDGE MA**

**Workshop co-chairs:**

David C. Parkes, Harvard University, and Rakesh V. Vohra, University of Pennsylvania.

**Workshop participants:**

Rediet Abebe, Ken Calvert, Elisa Celis, Yiling Chen, Bo Cowgill, Khari Douglas, Michael Ekstrand, Sharad Goel, Lily Hu, Ayelet Israeli, Chris Jung, Sampath Kannan, Dan Knoepfle, Hima Lakkaraju, Karen Levy, Katrina Ligett, Michael Luca, Eric Mibuari, Mallesh Pai, David C. Parkes, John Roemer, Aaron Roth, Ronitt Rubinfeld, Dhruv Sharma, Megan Stevenson, Prasanna Tambe, Berk Ustun, Suresh Venkatasubramanian, Rakesh Vohra, Hao Wang, Seth Weinberg, and Lindsey Zuloaga.

**Task force members:**

Liz Bradley (co-chair, University of Colorado, Boulder) Sampath Kannan (co-chair, University of Pennsylvania), Ronitt Rubinfeld (Massachusetts Institute of Technology), David C. Parkes (Harvard University), and Suresh Venkatasubramanian (University of Utah).

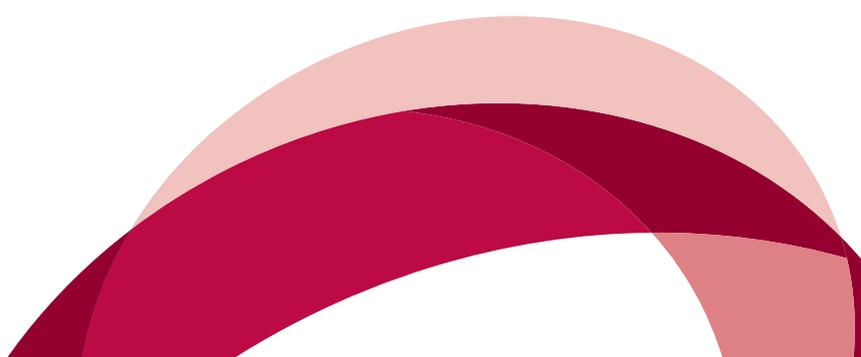

# 1. Overview

Algorithmic systems have been used to inform consequential decisions for at least a century. Recidivism prediction dates back to the 1920s (Burgess 1928), (Monachesi 1950), (Trainor 2015). Automated credit scoring dates began in the middle of the last century (McCorkell 2002), (Trainor 2015), (Lauer 2017), but the last decade has witnessed an acceleration in the adoption of prediction algorithms.

They are deployed to screen job applicants (Cowgill 2018a), (Cappelli, Tambe, and Yakubovich 2019) for the recommendation of products, people, and content, as well as in medicine (diagnostics and decision aids) (Ustun and Rudin 2017), ("MDCalc - Medical Calculators, Equations, Scores, and Guidelines" n.d.), criminal justice (Cowgill 2018b), (Megan Stevenson 2018) (setting bail and sentencing), facial recognition (Eubanks 2018), (Buolamwini and Gebru 2018), (Raji and Buolamwini 2019), lending and insurance (Jeong 2019), and the allocation of public services (Eubanks 2018), (Abebe and Goldner 2018).

The prominence of algorithmic methods has led to concerns regarding their systematic unfairness in their treatment of those whose behavior they are predicting. These concerns have found their way into the popular imagination through news accounts (Dastin 2018) and general interest books (O'Neill 2016), (Broussard 2018), (Noble 2018). Even when these algorithms are deployed in domains subject to regulation, it appears that existing regulation is poorly equipped to deal with this issue (Sullivan and Schweikart 2019).

The word 'fairness' in this context is a placeholder for three related equity concerns. First, such algorithms may systematically discriminate against individuals with a common ethnicity, religion, or gender, irrespective of whether the relevant group enjoys legal protections. The second is that these algorithms fail to treat people as individuals. Third, who gets to decide how algorithms are designed and deployed. These concerns are present when humans, unaided, make predictions.

So what is new here? *Scale* for one. These algorithms are being implemented precisely so as to scale up the number of instances a human decision maker can handle. Recruiters, for example, can process thousands of resumes in the blink of an eye. As a consequence, errors that once might have been idiosyncratic become systematic. *Ubiquity*, is also novel — success in one context justifies usage in other domains. Credit scores, for example, are used in contexts well beyond what their inventors originally imagined. Thirdly, *accountability* must be considered. Who is responsible for an algorithm's predictions? How might one appeal against an algorithm? How does one ask an algorithm to consider additional information beyond what its designers already fixed upon?

The concern for fairness is often set up in competition with a concern for accuracy. The first is seen as difficult to measure and hard to pin down, not least because one is concerned with fairness along a variety of dimensions such as income, health, and access to opportunity. Measuring accuracy, on the other hand is seen as unambiguous and objective. Nothing could be farther from the truth. Decisions based on predictive models suffer from two kinds of errors that frequently move in opposite directions: false positives and false negatives. Further, the probability distribution over the two kinds of errors is not fixed but depends on the modeling choices of the designer. As a consequence, two different algorithms with identical false positive rates and false negative rates can make mistakes on very different sets of individuals with profound welfare consequences. Prediction also depends crucially on the availability of data and data can be compromised in many ways — unevenness of coverage, sample bias, and noise. Hence, there are no simple and portable takeaways.

Motivated by these considerations the CCC's Fairness and Accountability Task Force held a visioning workshop on May 22-23, 2019, that brought together computer science researchers with backgrounds in algorithmic decision making, machine learning, and data science with policy makers, legal experts, economists, and business leaders. The workshop discussed methods to ensure economic fairness in a data-driven world. Participants were asked to identify and frame what they thought were the most pressing issues and to outline some concrete problems. This document is a synthesis of these comments.

We begin with four broad remarks that are helpful to frame one's thinking. First is an **equity principle** for evaluating outcomes (see Roemer and Trannoy (2016)). Under this principle, outcomes such as educational attainment, health status, and employability are assumed to be determined by factors that are divided into two categories. The first, called *circumstances*, are factors beyond an individual's control, such as race, height, and social origin. The second, called *effort variables*, are factors for which individuals are assumed to be responsible. In practice, it can be challenging to distinguish between factors that constitute circumstance and factors that constitute effort. Under this principle, inequalities due to circumstances holding other factors fixed are viewed as unacceptable and therefore justify interventions. Inequalities that arise from efforts, holding circumstances fixed, may be considered acceptable.[4] A challenge is that it may not be possible to isolate 'effort' from circumstance, such as parental wealth.[5] Even were there a clear distinction between the two, circumstances can shape an individual's incentives to exert effort. Further, circumstances and efforts are not always observed, and unobserved efforts may be correlated with observed circumstances and observed efforts may be correlated with unobserved circumstances.

The second is a distinction between two kinds of discrimination: **taste-based** and **statistical**. To understand the difference, imagine a decision has to be made about some agent, say whether to give the agent a loan or give her a job. The decision maker sees information about the agent, including protected demographic information (gender, race etc). A decision maker who discriminates against an otherwise qualified agent as a matter of taste alone is said to exhibit taste-based discrimination. That is, the demographics of the agent directly affect the preferences of the decision maker (for instance, the decision maker finds working with people of a certain gender distasteful). In contrast, a decision maker who is unconcerned with the agent's demographics per se, but understands that the demographics are correlated with the fitness of the agent for the task at hand is said to exhibit statistical discrimination. Given imperfect information about the agent's fitness, the decision maker uses the demographic information to make statistically better decisions. In principle, statistical discrimination may vanish/attenuate if better information about the agent's fitness were available. These forms of discrimination are conceptually and legally different. Indeed, laws in the US do allow for certain forms of statistical discrimination (the burden is on the decision maker to prove that decisions made using only other information would be statistically worse). The distinction is important because understanding the source of discrimination informs the possible policy response. It is well understood since Becker (1957) that taste-based discrimination is attenuated by competition between decision makers with heterogeneity in taste. However, short of providing better information, policies to reduce statistical discrimination are less well understood.

The third relates to the burgeoning field of **fair machine learning** whose goal is to ensure that decisions guided by algorithms are equitable. Over the last several years, myriad formal definitions of fairness have been proposed and studied by the computer science community (Narayanan 2018), (Hutchinson and Mitchell 2019), (Mitchell et al. 2018). One idea calls for *similar* individuals to be treated *similarly* (Dwork et al. 2012), and requires an appropriate measure of similarity. Another idea calls for *group-based definitions*, requiring, for example, that algorithms have *approximately equal error rates across groups* defined by protected attributes, like race and gender (Calders and Verwer 2010), (Edwards and Storkey 2015), (Hardt et al. 2016), (Kamiran, Karim, and Zhang 2012), (Pedreshi, Ruggieri, and Turini 2008), (Zafar et al. 2015), (Zemel et al. 2013). However, Chouldechova (2017) and Kleinberg et al. (2018) show it is typically impossible to satisfy group-based constraints for different error measures simultaneously. Corbett-Davies et al. (2017) and Corbett-Davies and Goel (2018) further argue that group-based definitions have counterintuitive statistical properties and, in some cases, can harm the groups they were designed to protect. As a result, one might take a *process-based approach*, with decisions made by thresholding on an estimate of an individual's *risk* (e.g., risk of default on a loan, or risk of recidivism), in this sense holding all individuals to the same standard. But this thresholding approach generally violates formal individual- and group-based definitions of fairness. Lastly, in those settings where sensitive attributes can be used (e.g., in medicine), then preference-based notions of fairness and decoupled classifiers have been suggested, requiring, for example that one group does not "envy" the classifier used for another group (Zafar et al. 2017), (Dwork et al. 2018), (Ustun, Liu, and Parkes 2019). Others advocate for adopting a welfare-economics viewpoint in interpreting appeals to fairness (Hu and Chen 2019), (Mullainathan 2018).

---

[4] This principle is not immune to criticism. Some argue that individuals are entitled to benefit from their draw in the genetic lottery (circumstance), see Nozick (1974). Others, that equity requires all individuals be guaranteed a minimum level of welfare irrespective of circumstance or effort, see Rawls (1971).

[5] Hufe et al. (2017) for example resolves this difficulty by restricting attention to individuals who are very young. In this case, it is hard to argue that effort variables will play a significant role.

The fourth relates to **data biases** (Suresh and Guttag 2019). All statistical algorithms rely on training data, which implicitly encode the choices of algorithm designers and other decision makers. For example, facial recognition algorithms have been found to perform worse on dark-skinned individuals, in part because of a dearth of representative training data across subgroups (Buolamwini and Gebru 2018), (Raji and Buolamwini 2019). In other cases, the target of prediction (e.g., future arrest) is a poor — and potentially biased — proxy of the underlying act (e.g., conducting a crime). Finally, when the training data are themselves the product of ongoing algorithmic decisions, one can create feedback loops that reinforce historical inequities (Kallus and Zhou 2018), (Ensign et al. 2018), (Lum and Isaac 2016). Mitigating these biases in the data is arguably one of the most serious challenges facing the design of equitable algorithms.

## 2. Decision Making and Algorithms

At present, the technical literature focuses on 'fairness' at the algorithmic level. The algorithm's output, however, is but one among many inputs to a human decision maker. Therefore, unless the decision maker strictly follows the recommendation of the algorithm, any fairness requirements satisfied by the algorithm's output need not be satisfied by the actual decisions. Green and Chen (2019), for example, report on an mTurk study that shows participants were more likely to deviate upward from algorithmic risk assessments for black defendants. M. Stevenson and Doleac (2019) discuss the introduction of risk assessment in sentencing in Virginia, and document that only a subset of judges appeared to integrate algorithmic risk assessment in their decisions.

Even if an algorithm's output violates some measure of fairness, it need not follow that the final outcomes are worse than the status quo of decision making sans algorithmic support. Cowgill (2018a), for example, documents an instance where the introduction of algorithmic resume screening reduced discrimination against non-traditional candidates. Kleinberg et al. (2018) describe a policy simulation that suggests that risk assessments in conjunction with human decision making would lower racial disparities relative to judges deciding alone.

The discussion above suggests the following questions:

a) How do human decision makers interpret and integrate the output of algorithms?

b) When they deviate from the algorithmic recommendation is it in a systematic way?

c) What is the role of institutional incentives for decision makers?

d) What can one say about the design of an algorithm that results in fair (fairer?) decisions by the human, which complements human decision making?

e) What aspects of a decision process should be handled by an algorithm and what by a human to achieve desired outcomes?

f) The "insufficiently diverse research team" hypothesis, is often cited as a reason for unfair machine learning algorithms.[6] Yet, we have no systematic documentation of the effects of biased programmers or the effects of diverse AI workforce on the outputs created by practitioners (Whittaker et al. 2018).

## 3. Assessing Outcomes

The outcome of an intervention can differ from its predicted effect because of the existence of indirect effects or feedback loops; for example, see predictive policing (Lum and Isaac 2016), (Ensign et al. 2018), as well as bail (Cowgill 2018b) recommendation. Hence, in addition to good-faith guardrails based on expected effects, one should also monitor and evaluate outcomes. Thus, providing *ex ante* predictions is no less important than *ex post* evaluations for situations with feedback loops.

At present, there is a paucity of work that seeks to quantify the effect on outcomes across the many domains where we will see automated decision making.[7] Measuring the effect of an algorithm on an outcome is inherently difficult because decisions made (or influenced) by an algorithm may have happened identically in the absence of the algorithm. Randomized controlled trials would be a natural way to assess such effects, but randomization may be repugnant in some applications of interest and requires smart experimental design.[8] Short of randomized controlled trials, the *regression discontinuity method* ("RD" (D. S. Lee and Lemieux 2010)) is a useful tool for measuring causal impact. Many machine learning applications make use of a continuous prediction (or score) with a decision threshold for an intervention, and the RD method estimates causal effects by looking at examples slightly below and above this threshold (assuming they are otherwise essentially identical). Papers that have used RD to study the causal impact of algorithms include Cowgill (2018b), Berk (2017), Anderson et al. (2013), and M. Stevenson and Doleac (2019).

Another challenge is that the environments in which algorithm-assisted decision making are deployed are always in flux. Consider hiring — today a firm may value individuals with analytical skills but tomorrow people skills may become the priority. Automated tools for hiring may also lead to defining a more and more narrow set of characteristics to allow it to consider a larger set of candidates. See, for example, the advice given to job seekers here: https://www.jobscan.co/blog/top-resume-keywords-boost-resume/.

Metrics to measure the extent of discrimination sometimes play an important role in regulatory guidelines but are challenging to develop and tend to be narrow in scope with effects that are hard to anticipate. A first example is the "four-fifths rule" of EEOC guidelines, which states that if the selection rate for a protected group is less than four-fifths of that for the group with the highest rate then this constitutes evidence of adverse impact.[9] A second example is the use of a single metric to measure the performance of a system. Such a metric can easily miss inequality that arises from complex effects (Crenshaw 1989), (Grusky and Ku 2008), (Grusky and Ku 2008). The domains in which algorithms are deployed are highly complex and dynamic, and data can pick up intersectional and multi-dimensional sources of discrimination (Abebe, Kleinberg, and Weinberg 2019).

Strategic considerations also play a role. For any proposed metric, one needs to identify the affected parties and their possible responses.[10] Therefore, policies cannot be judged *ceteris paribus.* Some existing research has shown that changing the incentive structure of those implementing or using algorithmic recommendations can in itself also be a tool for change (see Kannan et al. (2017) for example).

Theoretical research, particularly "impossibility theorems" (Chouldechova 2017), (Kleinberg et al. 2018), reveal that multiple attractive fairness properties are impossible to achieve simultaneously. Hence, it is inevitable that someone's notion of fairness will be violated and that tradeoffs need to be made about what to prioritize. These results do not negate the need for improved algorithms. On the contrary, they underscore the need for informed discussion about fairness criteria and algorithmic approaches that are tailored to a given domain. Also, we must recognize that these impossibility results are not about algorithms, *per se*. Rather, they describe a feature of any decision process, including one that is executed entirely by humans.

---

[6] The implicit suggestion of the work of Buolamwini and Gebru (2018) on biases in facial recognition technology (FRT) is that were there more programmers with dark skin this wouldn't have happened.

[7] Some exceptions include work on discrimination and bias in the context of facial recognition technology (Buolamwini and Gebru 2018), online ads (Sweeney 2013), word-embeddings (Bolukbasi et al. 2016), search engines (Noble 2018) and health information (Abebe et al. 2019).

[8] In credit scoring, for example, Kilbertus et al. (2019) suggest approving everyone with a high enough score, and randomly approving applicants with an insufficient score.

[9] This can be applied to any decisions related to employees — including hiring, promotion, and training.

[10] For instance, there is a literature (Coate and Loury 1993) in economics on what role a community's belief that they will be treated fairly (such as in education or access or the job market) affects their incentives to invest in human capital.

The discussion above suggests the following questions:

a) How do existing standards (e.g., disparate impact standards for hiring or housing) affect participation decisions and other quantities that are not directly scrutinized?

b) In regard to endogenous algorithm bias (Cowgill 2018a), can we identify the interventions that could change or reduce it?

c) Can we usefully model the feedback loop when designing metrics, and can we understand when a deployed system will still be able to be used for inference on cause and effect?

d) How can we design automated systems that will do appropriate exploration in order to provide robust performance in changing environments?

e) Can we understand the common issues that prevent the adoption of algorithmic decision-making systems across domains and the common issues that produce harm across multiple domains?

## 4. Regulation and Monitoring

Prohibitions against discrimination in lending, housing, and hiring are codified in law but do not provide the precise way in which compliance will be monitored. Poorly designed regulations have costly consequences in terms of compliance costs for firms, as well as generate harm to individuals.

Some have argued for "output" regulation. The "four-fifths rule", mentioned above, is an example. Others favor "input" regulation because they are more easily monitored than outputs.

Another challenge is that the disruption of traditional organizational forms by platforms (e.g., taxis, hotels, headhunting firms) has dispersed decision making. Who is responsible for ensuring compliance on these platforms, and how can this be achieved? On the one hand, platforms may be immune to existing regulation. For example, Section 230 of the Communications Decency Act (CDA) protects online software platforms from the actions of its users. At the same time, litigation and investigations have yielded penalties and changes (Levy n.d.). Platforms may also enable visibility into (and oversight of) discrimination that was previously difficult to observe.

Platforms lower the transaction costs of search and matching. Some, such as Lyft, make the match. Others, such as AirBnB, assist in search by curating and organizing the relevant information and making recommendations.

Ostensibly innocuous, such recommendation and rating systems can have huge impacts. One area of concern, for example, is whether these kinds of systems can lead to the consumption of less diverse content. Although the effect of recommender systems on diversity is debated (Nguyen et al. 2014), (Fleder and Hosanagar 2009), (Möller et al. 2018), this would then mean algorithms having a role in creating filter bubbles, which have in turn been argued to exacerbate polarization.[11] All this raises a number of questions. What does informed consent mean? Who gets to decide what an individual sees?

Effective regulation requires the ability to observe the behavior of algorithmic systems, including decentralized systems involving algorithms and people. To see the entire machine learning pipeline facilitates evaluation, improvement (including "de-biasing"), and auditing. On the other hand, this kind of transparency can conflict with privacy considerations, hinder innovation, and otherwise change behavior.

---

[11] Although there is evidence that the connection between the internet and social media and polarization is weak, see Boxell, Gentzkow, and Shapiro (2017).

The discussion above suggests the following questions:

a) When is output regulation preferable to input regulation and vice-versa?

b) Does the regulation of algorithms result in firms abandoning algorithms in favor of less inspectable forms of decision-making?

c) If regulating inputs, which portion of the machine learning pipeline should be regulated?

d) How should platforms design the information and choices they offer their users to reduce discrimination? (Agan and Starr 2018)

e) How should recommenders or similar systems be designed to provide users with more control? (e.g. Ekstrand and Willemsen (2016), Yang et al. (2019)).

## 5. Educational and Workforce Implications

Is the human capital necessary to think carefully about fairness considerations as they relate to algorithmic systems in abundance? *What should judges know about machine learning and statistics? What should software engineers learn about ethical implications of their technologies in various applications?* There are also implications for the interdisciplinarity of experts needed to guide this issue (e.g., in setting a research agenda). *What is the relationship between domain and technical expertise in thinking about these issues? How should domain expertise and technical expertise be organized: within the same person or across several different experts?* How do we train computer scientists to understand and engage with broader contexts, and to communicate and engage with relevant experts to broaden this understanding? The prior literature on these questions related to training and ensuring a well-equipped workforce includes Deming and Noray (2018) on STEM careers and technological change, and Oostendorp (2019) and Colson (2019) in regard to data science training.

Looking forward, it seems important to understand the effect of different kinds of training on how well people will interact with AI based decisions, as well as understand the management and governance structure for AI decisions. *Are managers (or judges) who have some technical training more likely to use machine learning-based recommendations? Are they more or less likely to benefit from machine learning-based recommendations?* In regard to governance, *what is the appeals process and is there a role for 'AI councils'?* As our curriculum changes, we should also seek to understand *whether explicitly embedding ethics training for computer science students influences bias-related outcomes. What about labor outcomes? For example, does domain expertise help in data science careers or vice versa?*

## 6. Algorithm Research

Algorithm design is a huge, well-established community in computer science, with lots of great problem-solvers who would love to work on impactful problems. At the same time, fairness questions are inherently complex and multifaceted and incredibly important to get right.

Today, it is reasonable to posit that a lot of work is happening around the various concrete definitions that have been proposed — even though practitioners may find some or even much of this theoretical algorithmic work misguided.

Given that it is hard to understand the intent behind different formalisms, a challenge that this presents to algorithm designers is that it makes it difficult to identify the most promising, technical algorithmic problems on which to work. This raises the question of how to promote cross-field conversations so that researchers with both domain (moral philosophy, economics, sociology, legal scholarship) and some technical expertise can help others to find the right way to think about different properties, and even identify if there are still dozens of properties whose desirability is not unanimously agreed upon.

Suppose an algorithms researcher comes up with a new algorithm and proves that it achieves a technical property, say that it equalizes false-positives. What sanity checks should be executed to see if this is for a silly reason? To draw an analogue: in the context of algorithmic game theory, it wouldn't be interesting to design a protocol where honest behavior is a Nash equilibrium just because behavior doesn't affect payoff in any way. Might it be possible to develop a sense of what is necessary for a result to be interesting in the context of fairness and economics? Can we see a path towards a community of algorithm designers who also have enough domain expertise that they are capable of identifying promising new technical directions, and work that will be appreciated by domain experts?

## 7. Broader Considerations

Some discussion amongst participants went to concerns about academic credit and how the status quo may guide away from applied work, noting also that the context of more applied work can be helpful in attracting more diverse students into computer science (Whittaker et al. 2018).[12] Others asked how one might promote more engagement with social science and researchers with domain expertise as well as policy-makers.[13] There are also difficult ethical challenges with conducting empirical, data-driven research, as considered within the NSF supported PERVADE (Pervasive Data Ethics for Computational Research) project.[14]

A thread that ran through all the discussions at the meeting was a sense that the research community may 'narrow frame' the issues under consideration. This is the tendency to define the choices under review too narrowly. For example, the problem of selecting from applicants those most qualified to perform a certain function is not the same as guaranteeing that the applicant pool includes those who might otherwise be too disadvantaged to compete.

The focus on prediction also leads to narrow framing. Predicting the likelihood of showing up for a bail hearing is not the same as understanding the reasons why an individual may be a no-show. The focus on one as opposed to the other leads to different interventions that could have dramatically different impacts. Prediction in this context, generally leads to the question of whether the individual should be released or not. Understanding the reasons behind a no-show may suggest interventions that lower the barriers to individuals to showing up.

---

[12] There is also a role for organizations such as Black in AI in fostering the involvement of individuals from under-represented groups and advocating for taking a multi-disciplinary perspective in AI fairness.

[13] Some early examples are the FAT* conference (an explicit call for interdisciplinarity) and the many interdisciplinary, but non-archival workshops, such as the MD4SG workshop, which provides space for "problem pitches" as well as more traditional formats.

[14] This project is working on metrics for assessing and moderating risks to data subjects, discovering how existing ethical codes can be adapted and adopted, and disseminating evidence-based best practices for research ethics.

## 9. Workshop Participants:

| First Name | Last Name | Affiliation |
|---|---|---|
| Rediet | Abebe | Harvard University |
| Ken | Calvert | National Science Foundation |
| Elisa | Celis | Yale University |
| Yiling | Chen | Harvard University |
| Sandra | Corbett | CRA/CCC |
| Bo | Cowgill | Columbia University |
| Khari | Douglas | Computing Community Consortium |
| Michael | Ekstrand | Boise State University |
| Sharad | Goel | Stanford University |
| Lily | Hu | Harvard University |
| Ayelet | Israeli | Harvard University |
| Sabrina | Jacob | CRA/CCC |
| Chris | Jung | University of Pennsylvania |
| Sampath | Kannan | University of Pennsylvania / CCC |
| Dan | Knoepfle | Uber |
| Hima | Lakkaraju | Harvard University |
| Karen | Levy | Cornell University |
| Katrina | Ligett | Hebrew University of Jerusalem |
| Mike | Luca | Harvard University |
| Eric | Mibuari | Harvard University |
| Mallesh | Pai | Rice University |
| David | Parkes | Harvard University / CCC |
| John | Roemer | Yale University |
| Aaron | Roth | University of Pennsylvania |
| Ronitt | Rubinfeld | Massachusetts Institute of Technology / CCC |
| Dhruv | Sharma | FDIC |
| Megan | Stevenson | George Mason University |
| Prasanna | Tambe | University of Pennsylvania |
| Berk | Ustun | Harvard University |
| Suresh | Venkatasubramanian | University of Utah / CCC |
| Rakesh | Vohra | University of Pennsylvania |
| Hao | Wang | Harvard University |
| Seth | Weinberg | Princeton University |
| Lindsey | Zuloaga | Hirevue |
| Christina | Ilvento | Harvard University |
| Mark | Hill | University of Wisconsin-Madison |
| Dirk | Bergemann | Yale University |

# NOTES

# NOTES

# NOTES

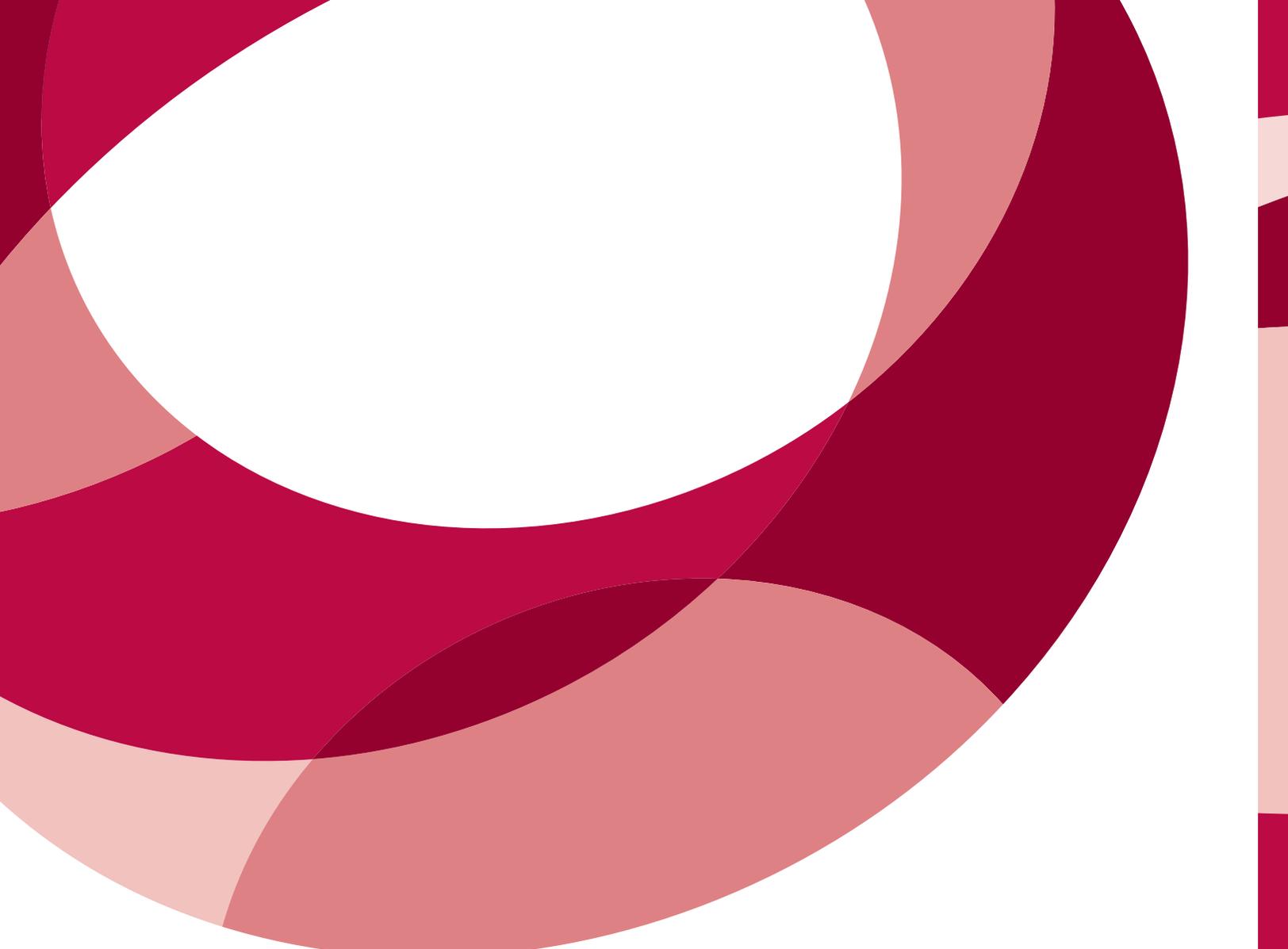

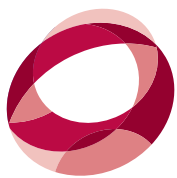

## CCC
**Computing Community Consortium**
Catalyst

1828 L Street, NW, Suite 800
Washington, DC 20036
P: 202 234 2111 F: 202 667 1066
www.cra.org cccinfo@cra.org